\documentclass[twocolumn,aps,preprintnumbers,amsmath,amssymb,showpacs]{revtex4}
\usepackage{subfig}
\usepackage{graphicx}
\usepackage{dcolumn}
\usepackage{bm}
\usepackage{color}
\usepackage{caption}
\begin{document}

\title{Stochastic Neutrino Mixing Mechanism}

\author{M. M. Guzzo$^1$} 
\author{P. C. de Holanda$^1$}
\author{O. L. G. Peres$^{1,2}$}
\author{E. M. Zavanin$^1$}
\affiliation{$^1$ Instituto de F\'\i sica
  Gleb Wataghin \\ Universidade Estadual de Campinas - UNICAMP \\
  Rua S\'ergio Buarque de Holanda, 777 \\
  13083-859 Campinas SP Brazil\\
$^2$Abdus Salam International Centre for Theoretical Physics, ICTP, I-34010, Trieste,Italy}
\date{\today}

\begin{abstract}
We propose a mechanism which provides an explanation of the Gallium and antineutrino reactor anomalies. Differently from original Pontecorvo's hypothesis, this mechanism is based on the phenomenological assumption in which the admixture of neutrino mass eigenstates in the moments of neutrino creation and detection can assume different configurations around the admixture parametrized by the usual values of the mixing angles $\theta_{12}$, $\theta_{23}$ and $\theta_{13}$. For simplicity, we assume a Gaussian distribution for the mixing angles in such a way that the average value of this distribution is given by the usual values of the mixing angles and the width of the Gaussian is denoted by $\alpha$. We show that the proposed mechanism provides a possible explanation for very short-baseline neutrino disappearance, necessary to accommodate Gallium and antineutrino reactor anomalies, which is not allowed in usual neutrino oscillations based on Pontecorvo's original hypotheses. We also can describe high-energy oscillation experiments, like LSND, Fermi and NuTeV, assuming a weakly energy dependent width parameter, $\alpha(E)$, that nicely fits all experimental results.
\end{abstract}

\maketitle

\section{Introduction}

Neutrino oscillations are a direct consequence of the assumption raised in the seminal article by Bruno Pontecorvo in 1957~\cite{Pontecorvo:1957cp} which asserts that neutrino states interacting with charged leptons through  weak interactions are  superpositions of neutrino states of non-vanishing definitive mass.  In his paper, Pontecorvo used an analogy with neutral kaon mixing to propose that neutrino-antineutrino transitions may occur. Although such matter-antimatter neutrino oscillation has not been observed, this idea formed the conceptual foundation for the quantitative theory of neutrino flavor oscillations, which were first developed by Maki, Nakagawa, and Sakata in 1962~\cite{Maki:1962mu} and further elaborated by several authors~\cite{Pontecorvo:1967fh,GribovPontecorvo,probability1,probability2,probability3}.

Under the assumption that flavor eigenstates are different superpositions of mass eigenstates, neutrino flavor oscillations can be described in the following way: as a neutrino propagates through space, the quantum mechanical phases of the mass eigenstates advance at slightly different rates due to the tiny differences in the neutrino mass eigenvalues. This results in a changing admixture of mass states as the neutrino travels. But a different admixture of mass states corresponds to a different flavor state. So a neutrino born as, say, an electron neutrino will be some different admixture of electron, muon, and tau neutrino after traveling some distance. Since the quantum mechanical phase advances in a periodic fashion, after some distance the state will return to the original admixture, and the neutrino will be again an electron neutrino. The electron flavor content of the neutrino will then continue to oscillate as long as the quantum mechanical state maintains coherence. It is because the mass differences between the neutrinos are small that the coherence length for neutrino oscillation is so long, making this microscopic quantum effect observable over macroscopic distances. 

Interesting enough, neutrino flavor oscillations are the basis of the so-called solutions of several puzzling neutrino observations recorded  along the last four decades. The solar neutrino deficit initially observed
in different experiments, counts on the neutrino oscillations resonantly enhanced by solar matter, the MSW phenomena~\cite{PhysRevD.17.2369,Mikheev:1986gs}, to explain the solar neutrino observations. Similarly the same oscillation parameters can explain the deficit in Kamland experiment. Also the strong zenith dependence of 
atmospheric neutrino and antineutrino data can be explained by evoking neutrino oscillations. Finally, 
 completely different sources of muon-neutrinos and muon-antineutrinos produced by meson decays in accelerators, confirms the necessity of neutrino oscillations to understand the observations~\cite{GonzalezGarcia:2007ib}. 
Furthermore, recent measurements of experiments collecting neutrinos from reactors observed the necessity of a nonvanshing neutrino mixing $\theta_{13}$~\cite{An:2012eh,Abe:2011fz}, composing a robust picture in favor of the Pontecorvo's hypotheses which give rise to neutrino oscillations. This complete scenario involving the three neutrino generations was recently analyzed in the Ref.~\cite{Tortola:2012te}.

On the other hand, one can argue that both Pontecorvo's hypotheses, i.e. neutrinos are massive and there exists neutrino mixing, have been experimentally confirmed only indirectly through their main consequence, precisely, the neutrino quantum oscillations. In fact, the first Pontecorvo's hypothesis which asserts that neutrinos are massive particles, has been directly tested in experiments involving precise measurements of the endpoint of the beta decay spectrum, if one is interested in the mass eigenstates present in what is called electron neutrino, or the kinematic behavior of the charged lepton produced in pion decay or tau decay, if one is interested in the mass eigenstates present in muon or tau neutrinos, respectively. Nevertheless, such observations had generated so far only superior limits to the values of the neutrino masses and no absolute values of such quantities were measured. 

The second Pontecorvo's hypothesis, i.e., the mixing hypothesis, could be directly tested carefully analyzing the composition of a neutrino beam just after its creation, very close to its source. An ideal experiment would consist of positioning a detector sensitive to different neutrino mass eigenstates very close to a neutrino source and observe its compositeness. Since detectors are only sensitive to neutrino flavor eingenstates, such an experiment cannot be realized. Nevertheless some hints on the neutrino compositeness could be achieved analyzing the flavor content of the recently created neutrino beam. The Pontecorvo mixing hypothesis foresee that close to their source neutrinos are found in a pure flavor eigenstate. Therefore, according to this Pontecorvo's hypotheses, very close to a reactor neutrinos have to be in pure electron flavor in the same way that very close to the pion decay pipe in an accelerator experiment, neutrinos have to be muon neutrinos or antineutrinos.

Nevertheless, there are indications that this is not always the case. Recent theoretical calculations of neutrino flux from nuclear reactors indicate that a larger  than previously expected neutrino flux is produced~\cite{Mention:2011rk,Huber:2011wv}. Such new fluxes are not entirely compatible with the short-baseline experiments which measure electron antineutrinos in distances of order 10 to 100 meters from nuclear reactors.   Furthermore, the procedure of calibration of the experiments GALLEX~\cite{Kaether201047}  and SAGE~\cite{PhysRevC.80.015807}  which measured neutrinos within distances as small as 1 m or so from the source raise also some incompatibility with  observations and predicted neutrino flux according to these new theoretical calculations. Such incompatibility has been called the anomaly of reactor antineutrino~\cite{Mention:2011rk,Huber:2011wv} and Gallium anomaly, respectively.  Several different phenomena have been evoked to explain  such anomalies~\cite{Mention:2011rk,PhysRevD.85.073012}.

In the present paper, we raise the possibility that the incompatibility of predictions and observations related to the reactor antineutrino and Gallium anomalies is a consequence of the usual interpretation of the Pontecorvo's quantum mixing hypothesis which defines, in a very fundamental and unique way, what is the admixture of mass eigenstates in a specific flavor neutrino eigenstate.  We will keep the usual interpretation that a neutrino produced in a reaction in which a charged lepton is involved is a neutrino of the same flavor of this charged lepton. Therefore, the antineutrino produced in a $\beta$-decay will be assumed to be of electron flavor. As well as, in a pion decay, once that a muon is involved, the corresponding neutrino will be assumed to be of muonic flavor and so on. Note that this is an arbitrary supposition once that neutrinos are not directed observed neither in creation nor in the detection processes. Nevertheless, different from the usual interpretation, we will assume that the admixture of mass eigenstates in the moment of neutrino creation is not unique but can vary for different neutrinos produced in that reaction. This implies also that what is called an electron neutrino in the creation moment can be a different combination of mass eigenstates from what is assumed to be an electron neutrino at the detection moment. Although unusual, we notice that such a hypothesis has never been tested so far and propitiates a possible explanation for the antineutrino reactor and Gallium anomalies, as we will see in the following. This new hypothesis and its consequences is what we call the Stochastic Neutrino Mixing Mechanism (SNMM). 


In order to appreciate how the SNMM works, we will analyze the particular case in which only two neutrinos are involved in the oscillation process. The generalization to the three neutrino case will be done in the next section. We propose relaxing the Pontecorvo's mixing hypothesis, allowing that neutrinos can be produced in an arbitrary superposition of neutrino mass eigenstates, each of them parametrized by a specific mixing angle $\theta_c$ in the creation moment:
\begin{equation}
\left|\nu_e^c\right\rangle = \cos\theta_c\left|\nu_1\right\rangle + \sin\theta_c\left|\nu_2\right\rangle,
\label{thetacreation}
\end{equation}
where $\theta_c$ can assume, in principle, any value in the interval $[0,\frac{\pi}{2}]$. The same assumption is made in the detection process, where the flavor state can also be identified in an arbitrary admixture of physical states, parametrized by a mixing angle at the detection moment in general different from the creation one, defined as $\theta_d$:
\begin{equation}
\left|\nu_e^d\right\rangle = \cos\theta_d\left|\nu_1\right\rangle + \sin\theta_d\left|\nu_2\right\rangle.
\label{thetadetection}
\end{equation}
Again, $\theta_d$ can assume any value in $[0,\frac{\pi}{2}]$. Under such assumption, after some distance $L$ from the source to the detector, the $\nu_e$ neutrino will present a survival probability calculated as
\begin{equation}
P^{one}_{\nu_e \rightarrow \nu_e} = \cos^2(\theta_d - \theta_c) - \sin2\theta_c\sin 2\theta_d \sin^2(\frac{\Delta m^2_{12} L}{4 E}),
\label{prob}
\end{equation}
where $E$  is the neutrino energy and $\Delta m^2_{12}$ is the usual squared mass difference between the mass eigenstates involved in the oscillation process.
Interesting enough, for the general case in which $\theta_c\neq\theta_d$, this survival probability can be smaller than the unity even in short baselines in which $L\rightarrow 0$. Such behavior, which is not allowed in usual oscillation processes, is the essence of the solution of the Gallium and reactor neutrino anomalies which will be explored in the next section in the more realistic case envolving three neutrinos.


\section{Three Neutrinos Case and the solution to the Gallium and Reactor Anomalies}

We propose relaxing the Pontecorvo's mixing hypothesis, allowing that each neutrino flavor eigenstate can be produced and detected in an arbitrary superposition of neutrino mass eigenstates around the usual admixture. 
In order to keep the success of neutrino oscillation observations, we assume that neutrinos are created and detected most of the time around the usual superposition of neutrino mass eigenstates which fit the oscillation phenomena, parametrized by the usual neutrino mixings $\sin^2 \theta_{12} = 0.320\pm 0.050$, $\sin^2 \theta_{23}=0.613^{+0.067}_{-0.247}$, and the recently measured $\sin^2 \theta_{13}= 0.025\pm 0.008$, in 3$\sigma$~\cite{Tortola:2012te}. In general, these specific angles are going to be assumed only as the averaged values of the actual mixing angles. Under this simple assumption, we will conclude that besides keeping the good fit of the observed long baseline neutrino oscillation phenomena, one can fit short baseline neutrino data setting a natural explanation for the anomaly of reactor antineutrino as well as Gallium anomalies. We assume, for simplicity, that such arbitrary superposition involves only the first two neutrino families. Therefore only variations around $\theta_{12}$ will be considered~\cite{comment}.

The $3\times 3$ mixing matrix at the moment of the neutrino creation ($U^c$) and at the detection moment ($U^d$) can be written as: 
\begin{equation}
\small
U^{c,d} =\left(
 \begin{array}{ccc}
  c^{c,d} c_{13} & -s^{c,d} c_{13} & s_{13} \\
  s^{c,d} c_{23} + c^{c,d} s_{23} s_{13} & c^{c,d} c_{23} - s^{c,d} s_{23} s_{13} & -s_{23} c_{13} \\
  s^{c,d} s_{23} - c^{c,d} c_{23} s_{13} & c^{c,d} s_{23} + s^{c,d} c_{23} s_{13}  & c_{23} c_{13}  \\
   \end{array}\right)
	\label{Ui}
\nonumber
\end{equation}
where $c_{ij}=\cos \theta_{ij}$, $s_{ij}=\sin\theta_{ij}$, $c^{c,d}=\cos\theta_{c,d}$ and $s^{c,d}=\sin\theta_{c,d}$, and $\theta_{c,d}$ can assume values in the interval $[0,\pi/2]$.

The one particle electron neutrino survival probability can be computed:
\begin{equation}
\small
P^{one}_{\nu_e\rightarrow \nu_e} = \left(\sum_{\gamma}{U^{c}_{1\gamma}U^{d}_{1\gamma}}\right)^2 - 4\sum_{\gamma>\beta}{U^{c}_{1\gamma}U^{d}_{1\gamma}U^{c}_{1\beta}U^{d}_{1\beta} \sin^2\left(\frac{\Delta m^2_{\gamma\beta} L}{4E}\right)}.
\end{equation}
where $\gamma$ and $\beta$ run from 1 to 3.
And then, averaging over different mixing angles, the total probability becomes:

\begin{equation}
P_{\nu_e \rightarrow \nu_e} = \int_0^{\pi/2} P_{\nu_e \rightarrow \nu_e}^{one} f(\theta_c)f(\theta_d)d\theta_c d\theta_d
\label{Preal3fam}
\end{equation}
where $f(\theta_c)$ and $f(\theta_d)$ are the distribution functions of the mixing angles involving only the electronic-muonic channel at the creation and detection instants, respectively. To keep the good fit of oscillation hypothesis with solar neutrino data and long baseline reactor observations, we choose these distribution functions as:
\begin{equation}
f(\theta_{c,d}) = \frac{1}{\sqrt{N_{c,d}}}e^{-(\frac{\theta_{c,d}-\theta_{12}}{\alpha_{c,d}})^2} ,
\end{equation}
which guarantees that mixing angles $\theta_{c,d}$ will present an
average value given by $\theta_{12}$.
In the above equation, $\alpha_{c,d}$ are the Gaussian widths at the creation and detection instants, respectively, and we will assume, for simplicity, $\alpha_c=\alpha_d=\alpha$. The normalization, $N_{c,d}$,  is computed by imposing $\int_{0}^{\frac{\pi}{2}} f(\theta_{c,d})d\theta_{c,d}~=~1$. Note that in the limit case when $\alpha\to 0$, we recover the usual Pontecorvo mechanism.

\begin{figure}
\centering
	\includegraphics[trim = 30mm 10mm 30mm 30mm, scale=.3]{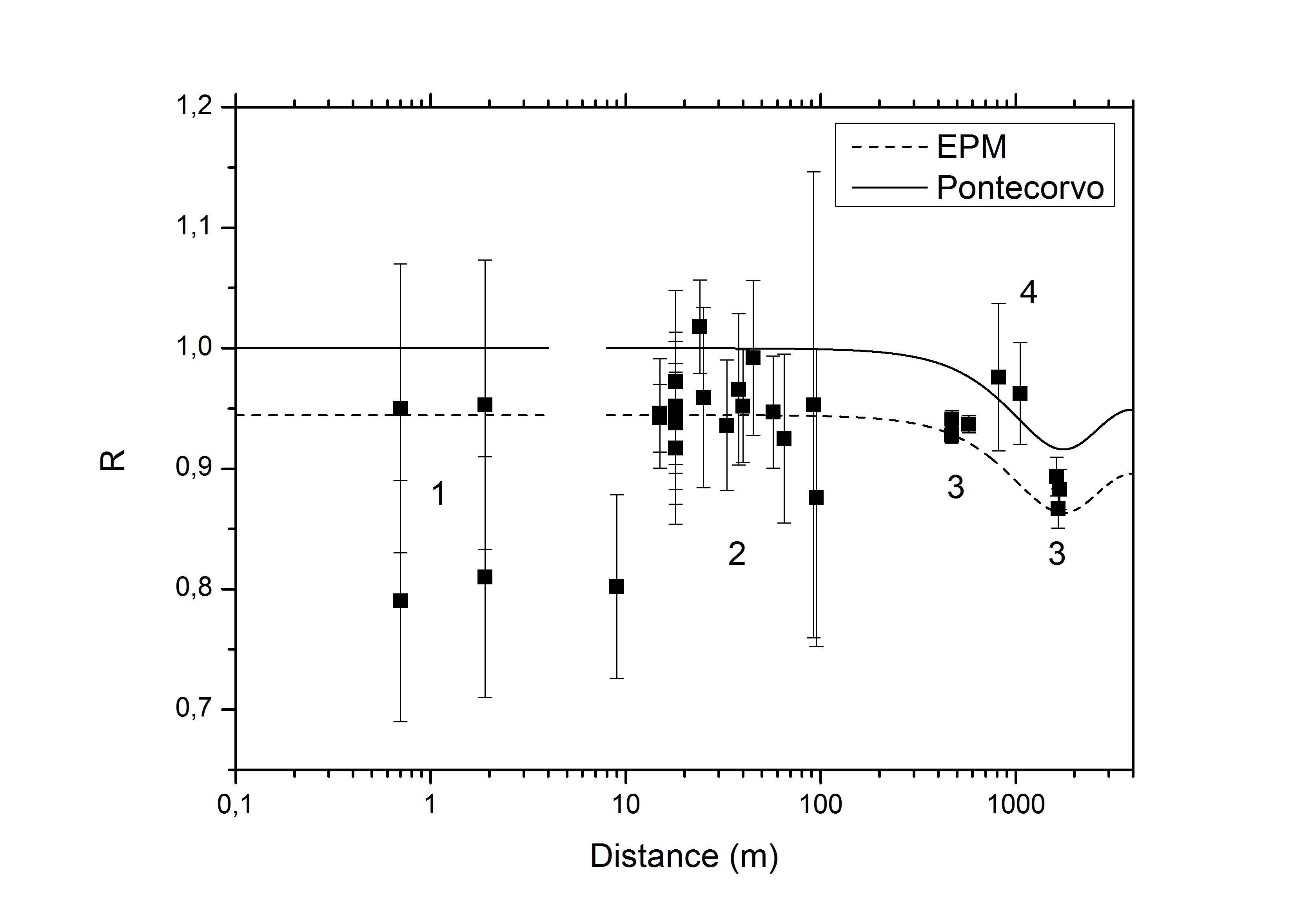}
		\caption{Comparison between the standard Pontecorvo's hypothesis prediction and the SNMM one plotted using the best fit value for the Gaussian width~$\alpha = 0.174$. The experimental points are distributed in the following way: 1. GALLEX and SAGE data;
2. old reactors~\cite{oldreactors};
3. Daya Bay data with free normalization;
4. Palo Verde and CHOOZ.}
	\label{fig:Grafico3familias}
\end{figure}

Using all data of GALLEX and SAGE experiments~\cite{Kaether201047,PhysRevC.80.015807} (see also Ref. \cite{PhysRevC.83.065504}), old reactors~\cite{oldreactors,Mention:2011rk} as well as  the Daya Bay data~\cite{An:2012eh} with  a free normalization found according to the new flux calculations for reactor experiments, we perform a global analysis through the $\chi^2$ method, defining:
\begin{eqnarray}
\chi^2 = \sum_{i,j=1}^4(\vec{R}^t-\vec{R}^e)_i^T W^{-1}_{ij} (\vec{R}^t-\vec{R}^e)_j, \nonumber
\label{chi}
\end{eqnarray}
where $i$ and $j$ correspond to each one of the four sets of experiments indicated by the labels appearing in Fig.~\ref{fig:Grafico3familias}: $i,j=1$ for GALLEX and SAGE, $i,j=2$ for old reactor experiments~\cite{oldreactors}, $i,j=3$ for Daya Bay and $i,j=4$ for Chooz and Palo Verde. $W_{ij}$ is the correlation matrix \cite{Mention:2011rk}, in which correlations between data coming from reactors described by $i,j= 2$ are taken into account, while no correlation among other data is assumed, and column vector $\vec{R^e}$ collects the experimental data while $\vec{R^t}$ the corresponding theoretical predictions for reactor and gallium experiments. For reactors one has:

\begin{equation}
R^t_{\rm reactor} = \dfrac{ \int{P_{\nu_e \rightarrow \nu_{e}} S(E) \sigma(E) dE}}{\int{S(E) \sigma(E) dE}},
\label{reac}
\end{equation}
where $S(E)$ is the energy neutrino spectrum which can be found in reference \cite{PhysRevC.83.054615} and $\sigma(E)$ is the cross section~\cite{Mention:2011rk}.
In GALLEX and SAGE radioactive calibration experiments, the reactions of electron capture produce neutrinos of fixed energies. This implies:
\begin{equation}
R^t_{\rm gallium} = \dfrac{\int dV L^{-2}\sum_i{(B.R.)_i \sigma_i P_{\nu_e \rightarrow \nu_{e}}(L,E_i)}}{\int dV L^{-2}\sum_i{(B.R.)_i \sigma_i }}
\label{gallium}
\end{equation}
and the branching ratio (B.R.), the cross section $(\sigma_i)$ and the detector specifications are found in Tables 1 and 2 of Ref.~\cite{PhysRevC.83.065504} and references therein.

The set of data includes 4 points from GALLEX and SAGE~\cite{PhysRevC.83.065504}, 21 from old reactors~\cite{Mention:2011rk} as well as 6 from Daya Bay~\cite{An:2012eh}. We obtain the best fit value for~$\alpha = 0.174$ varying in the intervals $[0.141,0.201]$, $[0.117,0.222]$ and $[0.067,0.249]$  
at 90, 95 and 99\% C.L., respectively. 
 
This probability fits the data with a minimum $\chi_{min}^2~=~39.08$ which can be compared with the one obtained from the usual Pontecorvo's hypothesis resulting $\chi^2= 48.24$, for $31-1$ degrees of freedom.  The best fit of SNMM as well as the fit coming from the usual Pontecorvo's hypothesis are shown in Fig.~\ref{fig:Grafico3familias} where it can be seen that the SNMM provides a possible explanation for short-baseline neutrino disappearance, something which is not allowed in usual neutrino oscillations based on Pontecorvo's original hypotheses.

Before introducing our conclusions, we will add a possible extension of the SNMM scenario. Up to now, we have discussed the relaxation of the Pontecorvo's hypotheses assuming a Gaussian distribution for the mixing angle $\theta_{12}$ characterized by a constant value of the corresponding Gaussian width $\alpha$. Here we will observe that the SNMM can nicely fit several experiments assuming an energy dependence of such width. This is a consequence of the fact that, differently from low energy short-baseline experiments, high energy short-baseline experiments do not present any appearance or disappearance neutrino phenomenon. In fact, the neutrino disappearance is more intense in GALLEX and SAGE $^{37}$Ar and $^{57}$Cr sources than in reactor experiments. In the first case the ratio $R$ is smaller than unity nearly 14\% \cite{PhysRevC.83.065504}and in reactors $R$ is lower than unity nearly 6\% \cite{Mention:2011rk}. Note however that the neutrino energy released in $^{37}$Ar and $^{57}$Cr sources have average value of 740~KeV while neutrino from reactors possess a wide range of energy with a peak in 3.6~MeV.
Similar fact occurs in accelerator experiments. LSND~\cite{PhysRevLett.81.1774} shows an excess of electronic neutrinos for energies about 30 MeV. MiniBooNE~\cite{MiniBooNECollaboration:arXiv1207.4809} searched in two channels $\nu_{\mu} \rightarrow \nu_e$ and $\bar{\nu}_{\mu} \rightarrow \bar{\nu}_e$ for oscillations. In the energy range of $200 < E /$MeV $< 1250$ was found signal of oscillation in both channels, however, data suggest that the excess of events decreases when the neutrino/antineutrino energies increase. The experiment described in Ref. \cite{PhysRevLett.47.1576}, which we refer to as Fermi, worked in a different scale of energy, with peak in 30~GeV, searching for oscillation in the $\nu_{\mu} \rightarrow \nu_e$ channel and did not report any signal of oscillations. The same happened in NuTeV experiment~\cite{PhysRevLett.89.011804}. Executed with an average energy of about 200~GeV, it did not find signal of oscillations in both channels $\nu_{\mu} \rightarrow \nu_e$ and $\bar{\nu}_{\mu} \rightarrow \bar{\nu}_e$. The only possible exception is the experiment KARMEN \cite{Eitel200289} that was executed with energies of about 15 MeV, lower than LSND. Although it did not find any compelling excess related to the background, its measurement was associated with large uncertainties. 

\begin{figure}
	\centering
		\includegraphics[trim = 30mm 10mm 30mm 30mm, scale=.3]{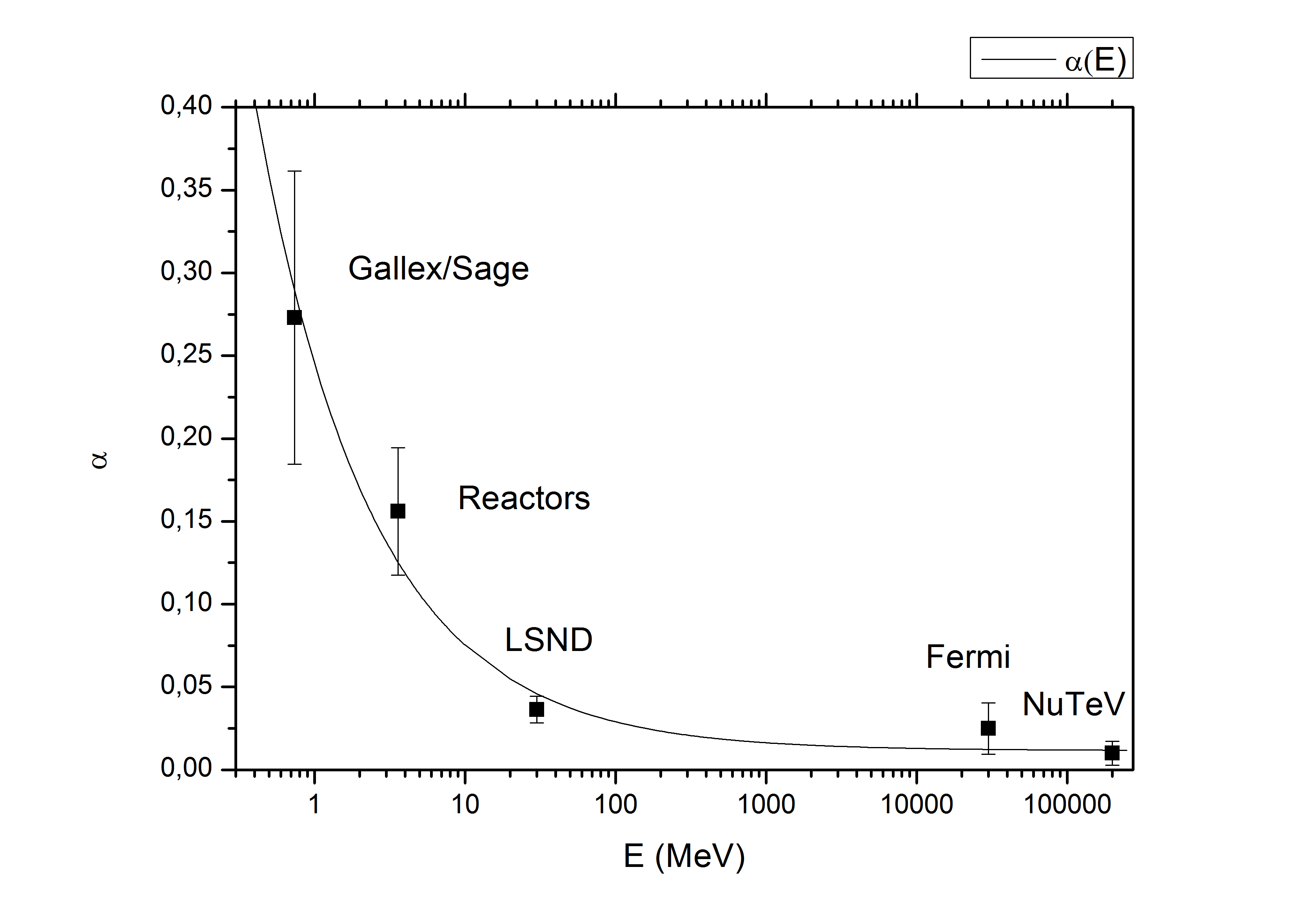}
		\caption{The Gaussian width $\alpha$ calculated to each set of experiments: GALLEX and SAGE, Reactors, LSND, Fermi and NuTeV. Points indicate the best fit at 68\%~C.L. and the curve shows the fitting to these points of the  functional form $\alpha(E) = A + (B/E)^n$, taking the best fit parameters 
		$A =~0.012$,~$B=~0.076 $ MeV and $n =$~0.565.}
	\label{fig:polalpha}
\end{figure}

The above cited experiments suggest that there is a relation between appearance/disappearance phenomena with the energy. Identifying this possible dependence, we independently calculate the free parameter $\alpha$ for each one of the following groups of experiments: 1. GALLEX and SAGE \cite{Kaether201047,PhysRevC.80.015807}, 2. all reactor data analyzed in \cite{Mention:2011rk} and Daya Bay \cite{An:2012eh}, 3. LSND \cite{PhysRevLett.81.1774}, 4. Fermi \cite{PhysRevLett.47.1576} and 5. NuTeV \cite{PhysRevLett.89.011804}, which result is indicated by the points and their uncertainties at 68\%~C.L.  in Fig.~\ref{fig:polalpha}. To fit  all data, we propose that the width have an energy dependence $\alpha(E) = A + (B/E)^n$. Taking the best fit parameters ($A =~0.012$,~$B=~0.076 $ MeV and $n =$~0.565), we also show this curve in Fig. \ref{fig:polalpha}.
\section{Conclusions and final comments}
The SNMM can accommodate data that indicate disappearance of electronic neutrinos/antineutrinos in very short-baseline experiments. Assuming that neutrino mixing angles can vary according to a Gaussian distribution around a preferable value given by the usual mixing angle $\theta_{12}$, a gaussian width $\alpha$ around $0.17$ can fit all experimental data in what is called Gallium and antineutrino reactor anomalies. Furthermore, identifying an energy dependence in short-baseline neutrino disappearance/appearance phenomena, we could explain different behaviors of high and low energy neutrino experiments assuming an energy dependence of the Gaussian width $\alpha$ which characterizes the SNMM.   

A few final comments are in order. First, we do not expect significant modifications of  previous analyzes involving solar, atmospheric and other long-baseline neutrino experiments due to the implementation of the SNMM. Only modifications of few percents in the initial neutrino flux predictions as well in the detection rate calculations in the analyses of those experiments can appear due to the SNMM. They can be accommodated in several uncertainties present in these analyses and will not substantially alter their results~\cite{comment}.

Secondly,  it is often assumed that contributions to neutrino masses come from new physics, while neutrino interactions are given by the Standard Model. Nevertheless, neutrinos are observed only indirectly through their interactions which produce charged leptons. This represents a challenge to implement realistic neutrino sectors in any model describing this particle. Some previous articles propose new approaches. An interesting discussion about the definition of a flavor  neutrino state and its relation with physical neutrino states can be found in Ref.~\cite{Grossman1995,GonzalezGarcia:2001mp,Bilenky:1992wv,Meloni:2009cg} in which possible mechanisms which can generate non trivial mixing matrices that can be different in the neutrino creation and in the neutrino detection are discussed. This is one of the requirements to implement the SNMM and can inspire the proposition of models which accomplish the mechanism.

We also propose possible tests to the SNMM. A muon neutrino detector located near a reactor could exclude this hypothesis in case no muon neutrino would be found. Such a detector could be based on muon neutrino elastic scattering on electrons, in a similar way which is discussed in Ref.~\cite{Adams:2008cm}, which is able to explore the muon neutrino at zero distance by $\nu_{\mu}e$ scattering.
Also,  
radioactive electron neutrino sources allocated inside experiments able to detect neutrinos through both charged and neutral currents channels (like as SNO~\cite{sno}) would test the SNMM hypothesis. A non-oscillation effect in the neutral current measurement and an oscillation effect in the charged current can favor SNMM in contrast to the sterile hypothesis, while a oscillation effect in the NC and CC measurement can indicate the presence of sterile neutrinos~\cite{PhysRevD.85.073012}.  

Finally, when we include an energy dependent gaussian width $\alpha$ we can fulfill the constraints on oscillation effects from low energy experiments, such as reactor and Gallium experiments, as well as the high energy experiments, such as FERMI, LSND, and NUTEV. We show in Fig.~2 that a weak energy dependence is sufficient to achieve a nice and consistent picture of SNMM as a solution to the reactor and Gallium anomalies.

\begin{acknowledgments}
The authors would like to thank FAPESP, CNPq and CAPES for several financial supports.
\end{acknowledgments}

%
%

\begin{thebibliography}{90}

\bibitem{Pontecorvo:1957cp}
B.~Pontecorvo, {\it Sov.Phys.JETP\/} {\bf 6}, 429 (1957).

\bibitem{Maki:1962mu}
Z.~Maki, M.~Nakagawa, S.~Sakata, {\it Prog.Theor.Phys.\/} {\bf 28}, 870 (1962).

\bibitem{Pontecorvo:1967fh}
B.~Pontecorvo, {\it Sov.Phys.JETP\/} {\bf 26}, 984 (1968).

\bibitem{GribovPontecorvo} V.~Gribov, B.~Pontecorvo, Phys. Lett. B 28 (1969)
493.

\bibitem{probability1} S.M.~Bilenky, B.~Pontecorvo, Sov. J. Nucl. Phys. 24 (1976) 316.

\bibitem{probability2} S.~Eliezer, A.R.~Swift, Nucl. Phys. B 105 (1976) 45;

\bibitem{probability3}H.~Fritzsch, P.~Minkowski, Phys. Lett. B62 (1976) 72.

\bibitem{PhysRevD.17.2369}
L.~Wolfenstein, {\it Phys. Rev. D\/} {\bf 17}, 2369 (1978).

\bibitem{Mikheev:1986gs}
S.~Mikheev, A.~Yu. Smirnov, {\it Sov.J.Nucl.Phys.\/} {\bf 42}, 913 (1985).

\bibitem{GonzalezGarcia:2007ib}
{For a review on neutrino oscillations and their phenomenological implications see:}
M. C.~Gonzalez-Garcia, M.~Maltoni, {\it Phys. Rept.\/} {\bf 460}, 1 (2008).

\bibitem{An:2012eh}
F.~An  {\it et~al.\/}, {\it Phys. Rev. Lett.\/} {\bf 108}, 171803 (2012).

\bibitem{Abe:2011fz}
Y.~Abe  {\it et~al.\/}, {\it Phys. Rev. Lett.\/} {\bf 108}, 131801 (2012).

\bibitem{Tortola:2012te}
D.~ V. Forero, M.~Tortola,  J.  W.F. ~Valle, {\it Phys. Rev.\/} {\bf D86}, 073012 (2012).

\bibitem{Mention:2011rk}
G.~Mention {\it et~al.\/}, {\it Phys. Rev.\/} {\bf D83}, 073006 (2011).

\bibitem{Huber:2011wv}
P.~Huber,   {\it Phys. Rev.\/} {\bf C84}, 024617 (2011).

\bibitem{Kaether201047}
F.~Kaether {\it et~al.\/}, {\it Phys. Lett. B\/} {\bf 685}, 47  (2010) and references therein.

\bibitem{PhysRevC.80.015807}
J.~N. Abdurashitov  {\it et~al.\/}, {\it Phys. Rev. C\/} {\bf 80}, 015807
  (2009) and references therein.


\bibitem{PhysRevD.85.073012}
P.A.N.~Machado, H.~Nunokawa, F.A.~Pereira dos Santos, R.Z.~Funchal, {\it Phys. Rev. D\/} {\bf 85}, 073012 (2012).

\bibitem{comment}
{A global analysis of short and long-baseline neutrino data including possible arbitrary variation of all mixing angles in the context of SNMM hypothesis 
   is in preparation by the authors}.

\bibitem{PhysRevC.83.065504}
C.~Giunti, M.~Laveder, {\it Phys. Rev. C\/} {\bf 83}, 065504 (2011).

\bibitem{oldreactors}
{The set of old reactor experiments included in this analysis is: ILL,
  Bugey-3/4, Rovno88-1S/3S, Rovno88-1I/2I, Rovno91, SRP-I, SRP-II, Rovno88-2S,
  Krasnoyarsk-I, Gosgen-, Bugey-3, Gosgen-II, Krasnoyarsk-III, Gosgen-III,
  Krasnoyarsk-II and Bugey-3. See \cite{Mention:2011rk} for details} .

\bibitem{PhysRevC.83.054615}
T.~A. Mueller {\it et~al.\/}, {\it Phys. Rev. C\/} {\bf 83}, 054615 (2011).


\bibitem{PhysRevLett.81.1774}
C.~Athanassopoulos {\it et~al.\/}, {\it Phys. Rev. Lett.\/} {\bf 81}, 1774
  (1998).

\bibitem{MiniBooNECollaboration:arXiv1207.4809}
A.~A. Aguilar-Arevalo  {\it et~al.\/}  arXiv:1207.4809 (2012).

\bibitem{PhysRevLett.47.1576}
N.~J. Baker  {\it et~al.\/}, {\it Phys. Rev. Lett.\/} {\bf 47}, 1576 (1981).

\bibitem{PhysRevLett.89.011804}
S.~Avvakumov  {\it et~al.\/}, {\it Phys. Rev. Lett.\/} {\bf 89}, 011804 (2002).

\bibitem{Eitel200289}
K.~Eitel, {\it Prog. in Part. and Nucl. Phys.\/} {\bf 48}, 89
  (2002).

\bibitem{Grossman1995}
Y. Grossman, {\it Phys. Lett. B \/} {\bf 359}, 141 (1995).
\bibitem{GonzalezGarcia:2001mp} 
  M.~C.~Gonzalez-Garcia, Y.~Grossman, A.~Gusso and Y.~Nir,
  Phys.\ Rev.\ D {\bf 64}, 096006 (2001)
  [hep-ph/0105159].
\bibitem{Bilenky:1992wv} 
  S.~M.~Bilenky and C.~Giunti,
  Phys.\ Lett.\ B {\bf 300}, 137 (1993)
  [hep-ph/9211269].
\bibitem{Meloni:2009cg} 
  D.~Meloni, T.~Ohlsson, W.~Winter and H.~Zhang,
  JHEP {\bf 1004}, 041 (2010)
  [arXiv:0912.2735 [hep-ph]].

\bibitem {sno}
Q. R. Ahmad {\it et~al.\/}, {\it Phys. Rev. Lett.\/} {\bf 89}, 011301 (2002).

\bibitem{Adams:2008cm}
T.~Adams {\it et al.}  [NuSOnG Collaboration],
Int.\ J.\ Mod.\ Phys.\ A {\bf 24}, 671 (2009)
  [arXiv:0803.0354 [hep-ph]].


\end{thebibliography}

\end{document}